# Dispersion Relations and Their Symmetry Properties for Electromagnetic and Elastic Metamaterials in Two Dimensions

Ying Wu and Zhao-Qing Zhang*

Department of Physics, The Hong Kong University of Science and Technology,

Clear Water Bay, Kowloon, Hong Kong, China

Abstract

In the framework of multiple-scattering theory, we show that the dispersion relations of certain electromagnetic (EM) and elastic metamaterials can be obtained analytically in the long-wavelength limit. Specific examples are given to the two-dimensional metamaterials with cylindrical inclusions arranged in square and triangular lattices. The role played by the lattice structure in determining whether a dispersion relation is isotropic or not is shown explicitly. Different lattice dependences between EM and elastic metamaterials are also shown. In the case of isotropic dispersions, our results coincide with those of isotropic effective medium theories obtained previously for EM and elastic metamaterials, respectively, and, therefore, provide a more fundamental support to those theories. In the case of elastic metamaterials with anisotropic dispersions, our analytical results can provide an anisotropic effective medium theory in the form of Christoffel's equation. In this case, the isotropic effective medium theory can describe accurately the angle-averaged dispersion relations. The properties of anisotropic dispersions are discussed and verified by numerical calculations of a realistic elastic metamaterial.

* To whom correspondence should be addressed. Email:phzzhang@ust.hk

# I  Introduction

In the past two decades, with the development of photonic and phononic crystals, much attention was paid to the wave propagation in periodic composites [1-4]. The dispersion relation of classical waves in a periodic structure is always linear, i.e., $\omega(\vec{k}) = c(\hat{k})k$, in the quasi-static limit, i.e., $\omega \to 0$, because the wave does not probe the fine structure of the lattice [2].

For electromagnetic (EM) waves propagating in photonic crystals, the effective speed, $\lim_{|\vec{k}| \to 0} c(\hat{k})$, can be isotropic, uniaxial or biaxial [5], which also means that the dispersion relations can be isotropic or anisotropic. This is because the dielectric constant, $\varepsilon_{ik}$, forms a tensor of rank two that can be determined by three independent quantities, known as the three principal values. Here, we consider only the case of isotropic inclusions and any anisotropy in the effective speed is caused by the lattice structures. Because of the symmetries of the lattice structure, the number of different principal values may be less than three. If all the principal values are the same, the effective speed is isotropic [6]. This normally requires that the unit cell of a photonic crystal possesses high symmetry in both its inclusion and lattice structure. Common structures like spherical inclusions arranged in *fcc*, *sc* and diamond lattices

in three-dimensions (3D) [5] and cylindrical inclusions in lattices with three-, four- and six-fold symmetries in two-dimensions (2D) are examples [7,8]. It should be noted that the isotropy in 2D case is only limited in the periodicity plane, say, the *xy*-plane.

For elastic wave propagating in elastic phononic crystals [9,10], the situation is different because elastic waves have two kinds of modes, i.e., longitudinal wave and transverse wave. Here, the term "elastic phononic crystal" is used when the host background is a solid material, which makes the whole system elastic for small stresses and the system can support both longitudinal and transverse waves. Since the elastic modulus tensor in a periodic structure is a tensor of rank four and has 21 different components in general [11, 12], the dispersion relations become much more complicated. However, the number of independent components can be greatly reduced if the lattice possesses certain symmetries. For example, there are three independent elastic moduli for a cubic lattice in 3D or a square lattice in 2D, i.e., one bulk modulus and two shear moduli; but only two independent elastic moduli for a triangular lattice in 2D, i.e., one bulk modulus and one shear modulus [11,12]. The latter case gives rise to an isotropic dispersion. Recently, effective speeds for certain elastic phononic crystals in both 2D and 3D have been studied by using the plane-wave-expansion (PWE) method [13,14]. It was found that effective speeds are anisotropic in general in 3D [13] . In 2D, the anisotropy is found to be distinct for the square lattice while very weak for the triangular structure [14]. The weak anisotropy is actually an artifact of the PWE method.

Recently, a great deal of attention has also been attracted by metamaterials, which can exhibit intriguing phenomena, such as negative refraction and superlensing effect, etc [15]. A metamaterial usually possesses low frequency resonances in its building blocks and its unusual physical properties can be found above the resonant frequencies [16]. For a metamaterial with a periodic lattice structure, the dispersion relations also become complicated due to the built-in resonances in its building blocks. The symmetry properties discussed previously can only apply to the lowest bands when $\omega \to 0$. However, for metamaterials, more interesting bands are those low-lying bands above the resonances. To the best of our knowledge, the symmetry properties of these bands have not been addressed before. Knowing these properties is both academically interesting and could be useful in applications. For example, it is interesting to know whether the symmetry properties of those low-lying bands remain the same as those of the lowest bands. The symmetry properties of those low-lying bands have direct effects on the wave propagation behavior through metamaterials. An anisotropic dispersion can have applications in self-collimation [17]. Wave propagation through metamaterials can be conveniently described by using effective parameters. In the case of isotropic dispersions, some effective medium theories (EMTs) have been proposed previously [18-20]. However, to the best of our knowledge an EMT for anisotropic dispersion relations is still lacking.

In this work, we use multiple-scattering theory (MST) [4, 18] to study the dispersion relations of EM and elastic metamaterials in 2D with cylindrical inclusions in the long-wavelength limit. The MST, in conjunction with a lattice sum, is known to produce exact dispersion relations for a periodic structure. Unlike the PWE method, the MST can treat the symmetry of a lattice structure rigorously. In the long-wavelength limit, MST can provide analytical dispersion relations for certain EM and elastic metamaterials. Here we derive the analytical dispersion relations for both triangular and square lattices. In the case of EM metamaterials, we show that both triangular and square lattices exhibit isotropic dispersion relations for all low-lying bands that satisfy the long-wavelength limit. In the case of elastic metamaterials, we show that the dispersion relations are isotropic for triangular lattices while anisotropic for square lattices with a higher anisotropy in the transverse branches. The role played by the lattice structure in determining the symmetry properties of dispersion relations is clearly revealed. Different lattice dependence between EM and elastic metamaterials is also shown explicitly. Our results show that the symmetry properties in the dispersion relations of the low-lying bands are the same as that in the lowest bands. In the case of isotropic dispersion, we also find that the EMTs obtained previously for EM metamaterials [12] and elastic metamaterials [19] in the long-wavelength limit produce exactly the same dispersion relations obtained here. Thus, our results also provide a more fundamental support to those EMTs. In the case of anisotropic dispersion, our analytical results can provide an anisotropic EMT in the form of Christoffel's equation. In the case of small anisotropy, we show that the isotropic EMT can still accurately predict the angle-averaged dispersion relations. This is verified numerically by using a realistic elastic metamaterial.

This paper is organized as follows. The multiple-scattering formalism for 2D EM metamaterial and the derivations of dispersion relations for triangular and square lattices are briefly presented in Section II. The corresponding derivations for elastic metamaterials and the anisotropic EMT in the form of Christoffel's equation are presented in Section III. Also, in Section III, the properties of anisotropic dispersion relations are discussed and verified by the numerical calculation of a realistic elastic metamaterial. Conclusions are drawn in Section IV.

## II. 2D Electromagnetic Metamaterials

The 2D EM metamaterials considered in this study are composed of parallel identical cylindrical inclusions arranged in a triangular or square lattice in the *xy*-plane embedded in an isotropic matrix. The cylindrical inclusions possess permittivity $\varepsilon_s$, permeability $\mu_s$ and radius $r_s$. The corresponding material parameters of the matrix are $\varepsilon_0$ and $\mu_0$, respectively. For a continuous wave incidence, the wave vectors in the scatterer and matrix are $k_s = \omega\sqrt{\varepsilon_s}\sqrt{\mu_s}$ and $k_0 = \omega\sqrt{\varepsilon_0}\sqrt{\mu_0}$, respectively, where $\omega$ is the angular frequency. In our study, we consider only the case of transverse magnetic (TM) waves, i.e., the electric filed is parallel to the cylinders' axes. The case of transverse electric (TE) waves can be obtained from the TM waves by interchanging the roles of $\varepsilon$ and $\mu$ due to the symmetry of Maxwell's Equations.

The dispersion relations of such an EM metamaterial can be obtained by solving a set of self-consistent equations derived from the MST. A brief derivation is given in Appendix A. The dispersion relations are determined by the following secular equation, i.e., Eq. (A.11):

$$\det\left|\sum_{m'} t_{mm'} \sum_{q \neq p} g_{m'm''} e^{i\vec{K} \cdot \vec{R}_q} - \delta_{m'm''}\right| = 0, \quad (1)$$

where $t_{mm'} = D_m \delta_{mm'}$ represents the scattering matrix and $D_m$ is the Mie scattering coefficient given by Eq. (A.4) in Appendix A. The term $\sum_{q \neq p} g_{m'm''} e^{i\vec{K} \cdot \vec{R}_q}$ represents the lattice sum and $\vec{K}$ is the Bloch wave vector, which is expressed by $(K, \phi_K)$ in polar coordinates. Both $D_m$ and $\sum_{q \neq p} g_{m'm''} e^{i\vec{K} \cdot \vec{R}_q}$ are functions of $\omega$. Thus, the solution of Eq. (1) gives the relation between $\omega$ and $\vec{K}$, i.e., dispersion relations.

To solve Eq. (1), we need to evaluate the lattice sum. There are a few systematic techniques for calculating it, such as direct sum and Ewald sum [21]. Here, we use the technique proposed by Chin *et al*. [22, 23], which expresses the lattice sum as absolutely converging series. Besides its efficiency, it also allows us to have some physical insight into the analytical properties of the lattice sum [22]. The detailed procedure of calculating the lattice sum is presented in Appendix B. The results are shown in Eqs. (B.6) and (B.7), i.e.,

$$\sum_{q \neq p} g_{m'm''} e^{i\vec{K}\cdot\vec{R}_q} = S(m'-m''), \qquad (2)$$

where

$$S(n) = \frac{1}{J_{n+1}(k_0 a)} \left( \frac{4i^{n+1} k_0}{\Omega} \sum_h \frac{J_{n+1}(Q_h a)}{Q_h (k_0^2 - Q_h^2)} e^{-in\phi_h} - \left( H_1^{(1)}(k_0 a) + \frac{2i}{\pi k_0 a} \right) \delta_{n,0} \right) \quad (n \geq 0). \quad (3)$$

$$S(-n) = -S^*(n)$$

Here, $a$ denotes the lattice constant, $\Omega$ is the volume of the unit cell and $(Q_h, \phi_h)$ stands for the vector $\vec{Q}_h = \vec{K} + \vec{K}_h$ in polar coordinates, and $\vec{K}_h$ is the reciprocal lattice vector. $J_{n+1}(x)$ and $H_1^{(1)}(x)$ are the Bessel and Hankel functions, respectively. Eq. (2) shows that the lattice sum depends only on the difference of the indices $m'$ and $m''$. Substituting Eq. (3) into Eq. (1), we can solve the secular equation to obtain the dispersion relations.

Eq. (1) can be solved analytically in the following two limits: the quasi-static limit and the long-wavelength limit. Here we use the following definitions given in Ref. [24]. The quasi-static limit requires that $Ka$, $k_0 a$ and $k_s r_s$ are all much smaller than unity. However, in the long-wavelength limit, one can relax the requirement of $k_s r_s \ll 1$. In fact, in a metamaterial, $k_s r_s$ is always greater than one at frequencies above its built-in resonances [25]. In these limits, we take appropriate approximations in Bessel and Hankel functions in the lattice sum $S(m'-m'')$ and scattering matrix $t_{mm'}$ so that the secular equation can be simplified and the analytical solution can be obtained.

In the long-wavelength limit, i.e., $Ka \ll 1$ and $k_0 a \ll 1$, we take $J_n(x) \cong \frac{x^n}{2^n n!}$,

and $H_1^{(1)}(x) \cong -\frac{2i}{\pi x}$ for $x = k_0 r_s$, $k_0 a$ and $Ka$. It can be shown that all the terms in Eq. (1) with angular quantum numbers greater than 1 ($|m| > 1$) are small compared to the terms with $|m| \leq 1$ and can be neglected so that Eq. (1) becomes

$$\det\left|\sum_{m'} t_{mm'} S(m'-m'') - \delta_{mm''}\right| = 0, \quad (m, m', m'' = -1, 0, 1). \tag{4}$$

In the approximation, the term $\left(H_1^{(1)}(k_0 a) + \frac{2i}{\pi k_0 a}\right)\delta_{n,0}$ in the lattice sum $S(n)$ turns to zero. Now, we separate the summation $\sum_h \frac{J_{n+1}(Q_h a)}{Q_h(k_0^2 - Q_h^2)} e^{-in\phi_h}$ in Eq. (3) into two terms: the first is $\vec{K}_h = 0$ or $\vec{Q}_h \cong \vec{K}$, the second is the sum of all other terms with $\vec{K}_h \neq 0$ or $\vec{Q}_h \cong \vec{K}_h$. Thus, the lattice sum takes the following expression

$$S(n) \cong \frac{4i^{n+1}}{\Omega}\left(\frac{K^n}{k_0^n(k_0^2 - K^2)} e^{-in\phi_K} + \frac{2^{n+1} k_0 (n+1)!}{(k_0 a)^{n+1}} \sum_{h(\vec{K}_h \neq 0)} \frac{J_{n+1}(K_h a)}{K_h(k_0^2 - K_h^2)} e^{-in\phi_{Kh}}\right) \; (0 \leq n \leq 2), \tag{5}$$

$$S(-n) = -S^*(n)$$

where $(K_h, \phi_{Kh})$ denotes the polar coordinates of $\vec{K}_h$ and we have used the approximation $\frac{J_{n+1}(Ka)}{K(k_0^2 - K^2)} e^{-in\phi_h} \cong \frac{(Ka)^{n+1}}{2^{n+1}(n+1)! K(k_0^2 - K^2)} e^{-in\phi_K}$ in the first term. Eq. (5) shows explicitly the dependence of the lattice sum on the lattice structure in terms of the summation of all non zero reciprocal lattice vectors, $\vec{K}_h$, of the lattice. Next, we will show how the lattice structure affects the lattice sum and the symmetry of dispersion relations.

For a 2D triangular lattice with a lattice constant $a$, $\vec{K}_h = h_i \frac{4\pi}{\sqrt{3}a}\hat{i} + h_j \frac{4\pi}{\sqrt{3}a}\left(\frac{1}{2}\hat{i} + \frac{\sqrt{3}}{2}\hat{j}\right); h_i, h_j \in Z$. Here, $\hat{i}$ and $\hat{j}$ represent the unit vectors along the $x$- and $y$-axes in the reciprocal space, respectively. When $n \neq 0$, the summation in the second term of Eq. (5) is zero due to the symmetry of a triangular lattice. This can be shown in the following way. If we arbitrarily choose one reciprocal lattice vector, $(K_{h1}, \phi_{Kh1})$, then we can always find the other five reciprocal lattice vectors at $(K_{h1}, \phi_{Kh1} + N\pi/3)$, $N = 1,2,3,4,5$ such that $\sum_{N=0}^{5} e^{-inN\pi/3} = 0$. Thus, the summation in Eq. (5) becomes zero after summing over all the non-zero $\vec{K}_h$. When $n = 0$, the second term in Eq. (5) no longer sums to zero as $e^{-in\phi_{Kh}} = 1$. However, this term is on the order of $\omega^0$ and can be ignored in the long-wavelength limit as the first term in Eq. (5) is on the order of $\omega^{-2}$. Thus, Eq. (5) is further reduced to:

$$S(n) \cong \frac{8 i^{n+1} K^n}{\sqrt{3} a^2 k_0^n (k_0^2 - K^2)} e^{-in\phi_K}, \quad 0 \leq n \leq 2, \tag{6}$$

where we have used $\Omega = \sqrt{3} a^2 / 2$.

Substituting Eq. (6) into Eq. (4), we obtain:

$$\det \begin{vmatrix} -\frac{\sqrt{3} a^2 (k_0^2 - K^2)}{8} + i D_1 & \frac{K}{k_0} D_1 & -\frac{iK^2}{k_0^2} D_1 \\ -\frac{K}{k_0} D_0 & -\frac{\sqrt{3} a^2 (k_0^2 - K^2)}{8} + i D_0 & \frac{K}{k_0} D_0 \\ -\frac{iK^2}{k_0^2} D_1 & -\frac{K}{k_0} D_1 & -\frac{\sqrt{3} a^2 (k_0^2 - K^2)}{8} + i D_1 \end{vmatrix} = 0. \tag{7}$$

Since the above determinant does not depend on $\phi_K$, the dispersion relation obtained from Eq. (7) is isotropic. It can be shown that Eq. (7) gives following solution:

$$K^2 = \frac{(-8 i D_0 + \sqrt{3} k_0^2 a^2)(-8 i D_1 + \sqrt{3} k_0^2 a^2)}{\sqrt{3} a^2 (8 i D_1 + \sqrt{3} k_0^2 a^2)}. \tag{8}$$

In the long wavelength limit, it can be shown that the Mie scattering coefficients take

the following forms:

$$D_0 \cong \frac{1}{4} i\pi k_0^2 r_s^2 \left( \frac{\tilde{\varepsilon}_s - \varepsilon_0}{\varepsilon_0} \right), \tag{9}$$

and

$$D_1 \cong \frac{1}{4} i\pi k_0^2 r_s^2 \frac{\tilde{\mu}_s - \mu_0}{\tilde{\mu}_s + \mu_0}, \tag{10}$$

where $\tilde{\varepsilon}_s = 2F(k_s r_s)\varepsilon_s / \left[1 + k_s^2 r_s^2 \ln(k_0 r_s) F(k_s r_s) \mu_0 / \mu_s \right]$ and $\tilde{\mu}_s = \mu_s F(k_s r_s)/(1 - F(k_s r_s))$ with $F(k_s r_s) = J_1(k_s r_s)/(k_s r_s J_0(k_s r_s))$. By substituting Eqs. (9) and (10) into Eq. (8), we obtain the following relation for the dispersion relation:

$$K^2 = \omega^2 \varepsilon_e \mu_e, \tag{11}$$

where $\varepsilon_e = \left( p\left( \frac{\tilde{\varepsilon}_s - \varepsilon_0}{\varepsilon_0} \right) + 1 \right)\varepsilon_0$ and $\mu_e = \left(1 + p\frac{\tilde{\mu}_s - \mu_0}{\tilde{\mu}_s + \mu_0}\right) / \left(1 - p\frac{\tilde{\mu}_s - \mu_0}{\tilde{\mu}_s + \mu_0}\right)\mu_0$ with $p = 2\pi r_s^2 / \sqrt{3} a^2$ as the filling ratio. Eq. (11) gives the dispersion relations of all the low-lying bands in the long-wavelength limit. It is interesting to point out that Eq. (11) is identical to the 2D version of the EMT developed in the long-wavelength limit [20, 25, 26], in which $\varepsilon_e$ and $\mu_e$ are, respectively, the effective permittivity and permeability of the effective medium. Thus, in the case of isotropic dispersions, we have reproduced the analytical results obtained here.

If the metamaterial possesses a square lattice structure, the reciprocal lattice vectors are $\vec{K}_h = h_i \frac{2\pi}{a} \hat{i} + h_j \frac{2\pi}{a} \hat{j}$; $h_i, h_j \in Z$. Similar to the triangular case, when $n \neq 0$, the summation in the second term of Eq. (5) is zero due to the symmetry of the square lattice; when $n = 0$, the first term is again two orders of magnitude greater than the second term. Hence, the lattice sum is almost the same as the triangular lattice, except that $\Omega = a^2$, i.e.,

$$S(n) \cong \frac{4i^{n+1}K^n}{a^2 k_0^n \left(k_0^2 - K^2\right)} e^{-in\phi_K}, \ 0 \leq n \leq 2. \tag{12}$$

Substitute Eq. (12) into Eq. (4) and solve for the root of the determinant for a given $K$, we obtain the same result as Eq. (11), except that the filling fraction in the effective parameters now becomes $p = \pi r_s^2 / a^2$.

From the above results, we find that, for a 2D EM metamatieral in a triangular or square lattice, the dispersion relations of all the low-lying bands are isotropic and can be predicted exactly by the EMT obtained in the long-wave length limit. It should be pointed out that the inclusion does not have to be a homogeneous scatterer. In general, it can be a multi-component scatterer in the form of layered structures [19]. In that case, the Mie scattering coefficient can be obtained by using standard transfer-matrix method. If we further apply the condition $k_s r_s \ll 1$ in the quasi-static limit to Eq. (11), the dispersion relation is reduced to: $K^2 = \omega^2 \varepsilon_e \mu_e$ with $\varepsilon_e = \left( p \left( \frac{\varepsilon_s - \varepsilon_0}{\varepsilon_0} \right) + 1 \right) \varepsilon_0$ and $\mu_e = \left( 1 + p \frac{\mu_s - \mu_0}{\mu_s + \mu_0} \right) \Big/ \left( 1 - p \frac{\mu_s - \mu_0}{\mu_s + \mu_0} \right) \mu_0$. This coincides exactly with the 2D version of the well-known Maxwell-Garnett (MG) theory [20]. The analytical results obtained here can explain the excellent agreements found previously between the PWE results and the MG theory in the dispersion relations of 3D photonic crystals in the quasi-static limit [5]. Since our approach is more fundamental than that of EMTs, our results also provide a firm support to the previous EMTs.

In the discussions above, we have neglected |m|=2 terms in the lattice sum, which are higher order terms in the long-wavelength limit. However, at higher frequencies, quadrupole excitations of cylinders will give rise to quadrupole bands that are not isotropic near the Γ point. This has been seen recently in plasmonic bands induced by electric quadrupole resonances in 3D photonic crystals consisting of plasmonic spheres [27].

### III. Elastic Metamaterials

The elastic metamaterial considered here is composed of identical cylindrical inclusions with bulk modulus $\kappa_s$, shear modulus $\mu_s$, mass density $\rho_s$, and radius $r_s$, arranged in a triangular or square lattice embedded in an isotropic host with material parameters $(\kappa_0, \mu_0, \rho_0)$. Due to the translational symmetry along the cylinders' axes, say the z-axis, the vibrations along the z-axis and in the xy-plane are decoupled. Here, the more complicated case of the xy-mode is studied [29].

In the long-wavelength limit, the dispersion relations of a 2D elastic metamaterial can also be obtained analytically by using the MST in conjunction with the lattice sum. The techniques used are similar to those in the EM case while the vector property of elastic waves makes the problem more complicated. A brief derivation is given in Appendix C.

Similar to Eq. (1), the dispersion relations of an elastic metamaterial are determined by the solutions to the following secular equation, i.e., Eq. (C.5):

$$\det\left|\begin{pmatrix} T^{ll}G^l & T^{lt}G^t \\ T^{tl}G^l & T^{tt}G^t \end{pmatrix} - I\right| = 0, \quad (13)$$

where $T^{\alpha\beta}$ are $m \times m$ matrices with matrix elements $t_{mm'}^{\alpha\beta} = D_m^{\alpha\beta}\delta_{mm'}$ and $D_m^{\alpha\beta}$ are the elastic Mie-like scattering coefficients. The explicit forms of $D_m^{\alpha\beta}$ can be found in Appendix A of Ref. [19]. $\alpha$ and $\beta$ take the values of $l$ and $t$, which denote, respectively, the longitudinal and transverse waves. The corresponding wave vectors inside the host are $k_{l0} = \omega\sqrt{(\kappa_0 + \mu_0)/\rho_0}$ and $k_{t0} = \omega\sqrt{\mu_0/\rho_0}$, respectively. In Eq. (13), $G^l$ and $G^t$ are $m \times m$ matrices, whose elements are given by the lattice sum $G_{m'm''}^{\beta} = \sum_{q \neq p} g_{m'm''}^{\beta} e^{i\vec{K}\cdot\vec{R}_q}$ ($\beta = l, t$). The lattice sum is given by Eqs. (C.6) and (C.7), i.e.,

$$\sum_{q \neq p} g_{m'm''}^{\beta} e^{i\vec{K}\cdot\vec{R}_q} = S(\beta, m'-m''), \quad (14)$$

where

$$S(\beta, n) = \frac{1}{J_{n+1}(k_{\beta 0}a)}\left(\frac{4i^{n+1}k_{\beta 0}}{\Omega}\sum_h \frac{J_{n+1}(Q_h a)}{Q_h(k_{\beta 0}^2 - Q_h^2)}e^{-in\phi_h} - \left(H_1^{(1)}(k_{\beta 0}a) + \frac{2i}{\pi k_{\beta 0}a}\right)\delta_{n,0}\right) \quad (n \geq 0). (15)$$

$$S(\beta, -n) = -S^*(\beta, n)$$

Here, $\vec{Q}_h = (Q_h, \phi_h)$ denotes the vector $\vec{K} + \vec{K}_h$ and $\vec{K} = (K, \phi_K)$ and $\vec{K}_h = (K_h, \phi_{Kh})$ represent, respectively, the Block wave vector and reciprocal lattice vector.

In the long-wavelength limit, i.e., $Ka \ll 1$ and $k_{\beta 0}a \ll 1$ ($\beta = l, t$), the leading terms in Eq. (13) are those with angular quantum numbers $|m| \leq 2$. Thus, we only need to consider the terms with $0 \leq n \leq 4$ in Eq. (15). This is different from the EM case, in which the $|m|=2$ terms are higher order terms than the terms with $m = \pm 1$ and 0. As we will see later that this important difference can make the symmetry property of dispersion relations of elastic metamaterials distinct from those of EM metamaterials.

Similar to the EM case, the lattice sum for an elastic metamaterial, i.e., Eq. (15), can be rewritten as,

$$S(\beta,n) \cong \frac{4i^{n+1}}{\Omega}\left( \frac{K^n}{k_{\beta 0}^n (k_{\beta 0}^2 - K^2)} e^{-in\phi_K} + \frac{k_{\beta 0} 2^{n+1}(n+1)!}{(k_{\beta 0} a)^{n+1}} \sum_{h(K_h \neq 0)} \frac{J_{n+1}(K_h a)}{K_h (k_{\beta 0}^2 - K_h^2)} e^{-in\phi_{Kh}} \right) (n \geq 0). (16)$$

$$S(\beta,-n) = -S^*(\beta,n)$$

**A. Triangular Lattice – Isotropic Dispersion Relations**

When the elastic metamaterial has a triangular lattice structure, the situation is similar to the EM metamaterial with triangular lattice case. When $n = 0$, the summation in the second term of Eq. (16) can be neglected as it is two orders of magnitudes smaller than the first term. When $n \neq 0$, summation in the second term of Eq. (16) cancels among themselves to zero due to the symmetry of triangular lattice structure. Thus, the lattice sum becomes :

$$S(\beta,n) \cong \frac{8i^{n+1} K^n}{\sqrt{3} a^2 k_{\beta 0}^n (k_{\beta 0}^2 - K^2)} e^{-in\phi_K}, \ 0 \leq n \leq 4. \tag{17}$$

Substituting Eq. (17) into Eq. (13), we find the following two roots:

$$\begin{aligned}\left(K_1^{tri}\right)^2 &= F_1(\tilde{D}_1^{ll}) F_2(\tilde{D}_2^{ll}) \\ \left(K_2^{tri}\right)^2 &= F_1(\tilde{D}_1^{ll}) F_3(\tilde{D}_2^{ll}, \tilde{D}_0^{ll})\end{aligned} \tag{18}$$

with

$$F_1(\tilde{D}_1^{ll}) = -\frac{8i\tilde{D}_1^{ll}(\kappa_0 + \mu_0) - \frac{\sqrt{3}}{2} a^2 \omega^2 \rho_0}{\frac{\sqrt{3}}{2} a^2} \ ,$$

$$F_2\left(\tilde{D}_2^{ll}\right) = -\frac{4\tilde{D}_2^{ll}(\kappa_0+\mu_0)(\kappa_0+2\mu_0)+i\frac{\sqrt{3}}{2}a^2\omega^2\mu_0\rho_0}{\mu_0\left(4\tilde{D}_2^{ll}\kappa_0(\kappa_0+\mu_0)-i\frac{\sqrt{3}}{2}a^2\omega^2\mu_0\rho_0\right)}, \quad (19)$$

$$F_3(\tilde{D}_2^{ll},\tilde{D}_0^{ll}) = -\frac{\left(4\tilde{D}_0^{ll}(\kappa_0+\mu_0)+i\frac{\sqrt{3}}{2}a^2\omega^2\rho_0\right)\left(4\tilde{D}_2^{ll}(\kappa_0+\mu_0)(\kappa_0+2\mu_0)+i\frac{\sqrt{3}}{2}a^2\omega^2\mu_0\rho_0\right)}{(\kappa_0+\mu_0)\left(32\tilde{D}_0^{ll}\tilde{D}_2^{ll}\mu_0(\kappa_0+\mu_0)^2-i\frac{\sqrt{3}}{2}a^2\omega^2\rho_0\left(4\tilde{D}_2^{ll}\kappa_0(\kappa_0+\mu_0)-i\frac{\sqrt{3}}{2}a^2\omega^2\mu_0\rho_0\right)\right)},$$

where $\tilde{D}_m^{ll}$ is obtained from the Mie-like scattering coefficient $D_m^{ll}$ by taking the long-wavelength approximation. Details can be found in the Appendix B of Ref. [19]. It should be mentioned that $\tilde{D}_m^{ll}$ are all imaginary so that both roots are real. Since the roots $\left(K_1^{tri}\right)^2$ and $\left(K_2^{tri}\right)^2$ given in Eq. (18) do not depend on $\phi_K$, this implies that all the low-lying bands for an elastic metamaterial in a triangular structure are isotropic near the $\Gamma$ point. After examining Eq. (18) carefully, we find that these two roots can be expressed as $\left(K_1^{tri}\right)^2 = \omega^2 \rho_e/\mu_e$ and $\left(K_2^{tri}\right)^2 = \omega^2 \rho_e/(\kappa_e+\mu_e)$, where $\rho_e$, $\kappa_e$ and $\mu_e$ satisfy the following equations,

$$\frac{(\kappa_0-\kappa_e)}{(\mu_0+\kappa_e)} = \frac{4\tilde{D}_0^{ll}}{i\Omega k_{l0}^2},$$

$$\frac{(\rho_0-\rho_e)}{\rho_0} = -\frac{8\tilde{D}_1^{ll}}{i\Omega k_{l0}^2}, \quad (20)$$

$$\frac{\mu_0(\mu_0-\mu_e)}{(\kappa_0\mu_0+(\kappa_0+2\mu_0)\mu_e)} = \frac{4\tilde{D}_2^{ll}}{i\Omega k_{l0}^2}.$$

Eq. (20) coincide exactly with the result of an EMT given in Ref. [19] for elastic metamaterials and $\rho_e$, $\kappa_e$ and $\mu_e$ are, respectively, the effective density, bulk modulus and shear modulus. Here we have demonstrated that the dispersion relations of all the low-lying bands of an elastic metamaterial with a triangular lattice structure is isotropic near the $\Gamma$-point and can be exactly reproduced by the EMT in the

## B. Square Lattice – Anisotropic Dispersion Relations and Christoffel's Equation

For the case of a square lattice, the lattice sum with $n=0$ is the same as that of the triangular lattice case. The second term of Eq. (16) is two orders smaller in magnitude than the first term so that can be neglected. When $n \neq 0$, the lattice sum is different from that of the triangular case. In this case, the reciprocal lattice vector of a square lattice is expressed by: $\vec{K}_h = h_i \frac{2\pi}{a} \hat{i} + h_j \frac{2\pi}{a} \hat{j}; \; h_i, h_j \in Z$. For an arbitrarily chosen reciprocal lattice vector $(K_{h1}, \phi_{Kh1})$, there always exist the other three at $(K_{h1}, \phi_{Kh1} + N\pi/2)$, $N = 1, 2, 3$, so that $\sum_{N=0}^{3} e^{-inN\pi/2} = 0$ when $1 \leq n \leq 3$ and $\sum_{N=0}^{3} e^{-inN\pi/2} = 4$ when $n = 4$. This indicates that the second term of Eq. (16) cancels to zero among themselves only when $1 \leq n \leq 3$. Thus, when $0 \leq n \leq 3$, the lattice sum can be written, in the long-wavelength limit, as

$$S(\beta, n) \cong \frac{4i^{n+1} K^n}{a^2 k_{\beta 0}^n (k_{\beta 0}^2 - K^2)} e^{-in\phi_K}, \; (0 \leq n \leq 3). \tag{21}$$

When $n = 4$, the second term of Eq. (16) does not vanish, leading to

$$S(\beta, 4) \cong \left( \frac{4iK^4}{a^2 k_{\beta 0}^4 (k_{\beta 0}^2 - K^2)} + \gamma_\beta e^{i4\phi_K} \right) e^{-i4\phi_K} \tag{22}$$

where

$$\gamma_\beta = \frac{16 \cdot 2^5 \cdot 5! i}{a^6 k_{\beta 0}^4} \left( \sum_{h_j=0}^{N} \sum_{h_i=1}^{N} \frac{J_5\left(2\pi \sqrt{h_i^2 + h_j^2}\right) e^{i4\arctan(h_j/h_i)}}{2\pi \sqrt{h_i^2 + h_j^2} \left( k_{\beta 0}^2 - (\frac{2\pi}{a})^2 (h_i^2 + h_j^2) \right)} \right). \tag{23}$$

It is the presence of the $\gamma_\beta$ term in Eq. (22) that makes the determinant in Eq. (13)

$\phi_K$-dependent, which, in turn, gives rise to anisotropic dispersion relations. The explicit expressions for $K_1^{squ}$ and $K_2^{squ}$ are very complicated. However, they can be expressed as the solutions of the following Christoffel's equation for an anisotropic medium [12], i.e.,

$$\det \begin{vmatrix} \cos^2\phi_K C_{11}\rho_e^{-1} + \sin^2\phi_K C_{66}\rho_e^{-1} - (\omega/K)^2 & \cos\phi_K \sin\phi_K (C_{12}+C_{66})\rho_e^{-1} \\ \cos\phi_K \sin\phi_K (C_{12}+C_{66})\rho_e^{-1} & \cos^2\phi_K C_{66}\rho_e^{-1} + \sin^2\phi_K C_{11}\rho_e^{-1} - (\omega/K)^2 \end{vmatrix} = 0,$$

(24)

where $\phi_K$ denotes angle between the Bloch wave vector $\vec{K}$ and the *x*-axis and $\rho_e$ is the effective density given in Eq. (20) with $\Omega = a^2$. The three effective moluli have the following expressions:

$$C_{11} = \kappa_e + \mu_e + \Delta_1,$$

$$C_{12} = \kappa_e - \mu_e - \Delta_1, \qquad (25)$$

and

$$C_{66} = \mu_e + \Delta_2,$$

where $\kappa_e$ and $\mu_e$ are given by Eq. (20) and $\Delta_{1,2} = \dfrac{\delta(\mu_0-\mu_e)^2}{\delta(\mu_0-\mu_e)\mp 8\rho_0}$, with $-$ and $+$ for $\Delta_1$ and $\Delta_2$ respectively. Here $\delta = i(k_{l0}^4\gamma_l - k_{t0}^4\gamma_t)a^2/\omega^2$ and $\gamma_{l,t}$ is given by Eq. (23). In the long-wavelength limit, it has the following expression:

$$\delta = -\frac{1920}{\pi^5}\frac{\kappa_0\rho_0}{\mu_0(\kappa_0+\mu_0)}\left(\sum_{h_j=0}^{N}\sum_{h_i=1}^{N}\frac{J_5\left(2\pi\sqrt{h_i^2+h_j^2}\right)e^{i4\arctan(h_j/h_i)}}{\left(\sqrt{h_i^2+h_j^2}\right)^5}\right). \qquad (26)$$

Eq. (24) gives the following two dispersion relations:

$$\left(\frac{\omega}{K_1}\right)^2 = \frac{1}{2\rho_e}\left(C_{11}+C_{66} - \sqrt{\sin^2(2\phi_K)(C_{12}+C_{66})^2 + \cos^2(2\phi_K)(C_{11}-C_{66})^2}\right),$$

$$\left(\frac{\omega}{K_2}\right)^2 = \frac{1}{2\rho_e}\left(C_{11}+C_{66} + \sqrt{\sin^2(2\phi_K)(C_{12}+C_{66})^2 + \cos^2(2\phi_K)(C_{11}-C_{66})^2}\right).$$

(27)

Since the origin of anisotropy comes from the term $\gamma_\beta \neq 0$, we expect that isotropy is recovered when $\gamma_\beta = 0$ (or $\delta = 0$). In this case, we have $C_{11} = C_{12} + 2C_{66}$ [12] and Eq. (27) gives the known wave speeds $(\omega/K_1)^2 = \mu_e/\rho_e$ and $(\omega/K_2)^2 = (\kappa_e + \mu_e)/\rho_e$ for transverse and longitudinal waves, respectively. For the case of anisotropic dispersions, Eq. (27) gives the dispersion relations for the quasi-transverse and quasi-longitudinal bands [12]. According to Eq. (27), $(\omega/K_1)^2$ has extrema at $\phi_K = 0$, and $\pi/4$, given by $(\mu_e + \Delta_2)/\rho_e$ and $(\mu_e + \Delta_1)/\rho_e$, respectively. Thus, $(\omega/K_2)^2$ oscillates between these two extrema. Similarly, $(\omega/K_2)^2$ oscillates between its two extrema, $(\kappa_e + \mu_e + \Delta_1)/\rho_e$ and $(\kappa_e + \mu_e + \Delta_2)/\rho_e$. If both $|\Delta_1|$ and $|\Delta_2|$ are much smaller than $|\mu_e|$ and $|\kappa_e + \mu_e|$, and the amplitude of the oscillation, $|\Delta_1 - \Delta_2|$, is small, the angle averaged dispersions, $(\omega/K_1)^2$ and $(\omega/K_2)^2$, can be well approximated by $\mu_e/\rho_e$ and $(\kappa_e + \mu_e)/\rho_e$, which are the results of isotropic EMT given by Eq. (20).

It should be pointed out that Eq. (27) describes the dispersion relations of all low-lying bands in the long-wavelength limit. In the quasi-static limit, $\kappa_e$ and $\mu_e$ appearing in Eq. (20) reduce to the following equations [19]

$$\frac{(\kappa_0 - \kappa_e)}{(\mu_0 + \kappa_e)} = \frac{\pi r_s^2}{a^2} \frac{(\kappa_0 - \kappa_s)}{(\mu_0 + \kappa_s)},$$
$$\frac{(\mu_0 - \mu_e)}{(\kappa_0 \mu_0 + (\kappa_0 + 2\mu_0)\mu_e)} = \frac{\pi r_s^2}{a^2} \frac{(\mu_0 - \mu_s)}{(\kappa_0 \mu_0 + (\kappa_0 + 2\mu_0)\mu_s)}. \tag{28}$$

Eqs. (25)-(28) give the analytical solutions for the anisotropic wave speeds for the case of square lattice. This result shows again that the symmetry properties of all the low-lying bands remain the same as that of the lowest bands.

In the following, we will study numerically the anisotropic properties found above by using a realistic metamaterial  The elastic metamaterial we study here is an array of silicone rubber cylinders embedded in an epoxy host in a square lattice [9, 19]. The silicone rubber in our system has density $\rho_s = 1.3 \times 10^3 kg/m^3$, longitudinal velocity $c_{ls} = \sqrt{(\kappa_s + \mu_s)/\rho_s} = 22.87 m/s$ and transverse velocity $c_{ts} = \sqrt{\mu_s/\rho_s} = 5.54 m/s$. The epoxy host possesses density $\rho_0 = 1.18 \times 10^3 kg/m^3$, longitudinal velocity $c_{l0} = \sqrt{(\kappa_0 + \mu_0)/\rho_0} = 2539.52 m/s$ and transverse velocity $c_{t0} = \sqrt{\mu_0/\rho_0} = 1160.80 m/s$. The radius of the scatter is $0.2a$, where $a$ is the lattice constant.  In the calculation, we set $a$, $\rho_0$ and $c_{t0}$ as unity for convenience.

Anisotropic dispersion relations can be easily seen from an equi-frequency surface (EFS). An EFS is a collection of all states in the $\vec{K}$ space that have the same frequency, say $\tilde{f}$, where dimensionless frequency, $\tilde{f} = f \frac{a}{c_{t0}} = \frac{\omega}{2\pi} \frac{a}{c_{t0}}$, is used. Fig. 1(a) shows two EFSs at $\tilde{f}_1 = 0.01$ with the inner one representing the quasi-longitudinal branch and the outer one the quasi-transverse branch.  Anisotropy is clearly seen in the quasi-transverse branch.  A weak anisotropy also exists in the quasi-longitudinal branch.  This can be seen more easily by plotting $Ka$ as a function of $\phi_K$ for each EFS.  The results are shown in Fig. 1(b) with open circles, which form two oscillating curves.  This implies that the $\gamma_\beta$ term in Eq. (22) induces oscillatory behavior of $Ka$ as a function of $\phi_K$. In this case, $\Delta_1 = 0.0157$ and $\Delta_2 = -0.0177$, which are very small compared to $\kappa_e$ (2.243) and $\mu_e$ (0.733).

From previous discussions, we know that at $\phi_K = 0$, $(\omega/K_1)^2$ and $(\omega/K_2)^2$ should reach their maximum and minimum respectively, which, in turn, implies $K_1$ and $K_2$ are at their minimum and maximum, respectively. This can be clearly seen from Fig. 1(b). For the upper branch, $Ka$ starts from its maximum of 0.0769 at $\phi_K = 0$, then bends down to its minimum of 0.0729 at $\phi_K = \pi/4$. For the lower branch, $Ka$ starts from its minimum of 0.0368 at $\phi_K = 0$, then turns up slightly to its maximum of 0.0374 at $\phi_K = \pi/4$. If we use the ratio $d_i = \left(K_i^{\max} - K_i^{\min}\right)/\langle K_i \rangle, (i=1,2)$ to characterize the amount of anisotropy, where $K_i^{\max}$, $K_i^{\min}$ and $\langle K_i \rangle$ are the maximum, minimum and average of $K_i$, we find $d_1 = 5.46\%$ and $\langle K_1 a \rangle = 0.0749$ for the transverse waves and $d_2 = 1.67\%$ and $\langle K_2 a \rangle = 0.0371$ for the longitudinal waves. The results obtained from Eq. (20), i.e., $K_t a = 0.0744$ and $K_l a = 0.0369$, are also plotted in Fig. 1 (b) in two horizontal solid lines. Fig. 1(b) demonstrates that the EMT can well predict the angle-averaged value of $K_i a$ in the case of anisotropy. Figs. 1(c) and 1(d) show another results calculated at $\tilde{f} = 0.03$, from which we see similar behavior of $Ka$ as a function of $\phi_K$. In this case, we find $d_1 = 5.39\%$ and $\langle K_1 a \rangle = 0.1603$ for the transverse waves and $d_2 = 1.20\%$ and $\langle K_2 a \rangle = 0.0793$ for the longitudinal waves. The EMT now gives $K_t a = 0.1599$ and $K_l a = 0.0794$, which are again close to the values of $\langle K_1 a \rangle$ and $\langle K_2 a \rangle$, respectively. The above results show that the dispersion relations predicted by the EMT can approximately reproduce the angle-averaged dispersion relations in the case of a square lattice. Fig. 1 only gives two examples at two frequencies, we also performed MST calculations for all low-lying bands. The angle-averaged dispersion relations are shown by circles in Fig. 2. Also plotted in Fig. 2 are the dispersion relations produced by the isotropic EMT for both quasi-longitudinal branches (solid curves) and quasi-transverse branches (dashed curves). The excellent agreements between two sets of results show clearly that the isotropic EMT is capable of predicting the angle-averaged dispersion relations for a square lattice in the case of small oscillations. It is also worth noting that although the condition of the long-wavelength limit requires $Ka \ll 1$, the analytical results so obtained still agree with the angle-averaged dispersion relations even when $Ka$ is as large as 0.3.

## V Conclusions

In the framework of MST, we have studied the dispersion relations of certain 2D EM and elastic metamaterials with cylindrical inclusions in the long-wavelength limit. For EM metamaterials, both triangular and square lattices exhibit isotropic dispersion relations in their low-lying bands. For elastic metamaterials, the dispersion relations of a triangular lattice are also isotropic while those of a square lattice are anisotropic with higher anisotropy for the transverse branch. The anisotropy is induced by the 4-fold symmetry of a square lattice and the terms whose angular quantum numbers are associated with $m = \pm 2$. For all the cases studied here, we find that the symmetry properties in the low-lying higher bands remain the same as their counterparts in the lowest bands. In the case of isotropic dispersions, our analytical results coincide with those of isotropic EMTs for EM and elastic metamaterials, which give a more fundamental support to those theories. In the case of anisotropic dispersions, our analytical results can provide an anisotropic EMT in the form of Christoffel's equation. In this case, the isotropic EMT can still accurately predict the angle-averaged dispersion relations when anisotropy is small. Although this work is done for 2D systems, we believe that the method shown here can also be used to study the dispersion relations of 3D EM and elastic metamaterials with spherical inclusions.

**Acknowledgement:**

The work was supported by Hong Kong RGC Grant No. 605008.

**Appendix A   Multiple-Scattering Formulism for EM Waves**

For TM waves, the electric field is parallel to the cylinder. For a sample with periodic structure, the solutions to the wave equations can be expanded by using Bessel functions, $J_m(x)$, and Hankel functions, $H_m^{(1)}(x)$. We first consider the scattering of a single scatterer (inclusion) by a continuous wave incidence. The incident wave on any scatterer, $p$, in general, has the form:

$$E_z^{in}(\vec{r}_p) = \sum_m a_m^p J_m(k_0 r_p) e^{im\theta_p} . \tag{A.1}$$

The corresponding scattered wave is expressed by:

$$E_z^{sca}(\vec{r}_p) = \sum_m b_m^p H_m^{(1)}(k_0 r_p) e^{im\theta_p} , \tag{A.2}$$

where $\vec{r} \equiv (r_p, \theta_p)$ are the polar coordinates originating at the center of the scatterer, $p$; $k_0 = \omega\sqrt{\varepsilon_0 \mu_0}$ is the wave vector in the matrix; $\omega$ is the angular frequency and $m$ represents the angular quantum number. The coefficients $b_m^p$ and $a_m^q$ are related through boundary conditions, which gives that:

$$b_m^p = \sum_{m'} t_{mm'} a_{m'}^p, \tag{A.3}$$

with $t_{mm'} = D_m \delta_{mm'}$. Here, $D_m$ is the Mie scattering coefficients of the scatterer, which can be expressed as [28]:

$$D_m = \frac{\mu_0 k_s J_m'(k_s r_s) J_m(k_0 r_s) - \mu_s k_0 J_m(k_s r_s) J_m'(k_0 r_s)}{\mu_s k_0 J_m(k_s r_s) H_m^{(1)}(k_0 r_s) - \mu_0 k_s J_m'(k_s r_s) H_m^{(1)}(k_0 r_s)}, \tag{A.4}$$

where $k_s = \omega \sqrt{\varepsilon_s \mu_s}$ is the wave vector in the scatterer.

For a collection of scatterers, the multiple-scattering between any of the two scatterers should be taken into account. Therefore, we also need to include in Eq. (A.1) the scattered waves coming from all other scatterers except for $p$ in a self-consistent way. This part can be written as:

$$\vec{u}_p^{in}(\vec{r}_p) = \sum_{q \neq p} \sum_{m''} b_{m''}^q H_{m''}^{(1)}(k_0 r_q) e^{im''\theta_q}. \tag{A.5}$$

where $(r_q, \theta_q)$ denote $\vec{r}_q$ in the polar coordinates originating at the center of scatterer $q$. Here, $\vec{r}_q$ and $\vec{r}_p$ refer to the same spatial point measured from the positions of scatterers $q$ and $p$, respectively. For simplicity, we choose the center of scatterer $p$ as the origin and denote the position of scatterer $q$ as $\vec{R}_q = (R_q, \Theta_q)$, which implies, $\vec{r}_q = \vec{r}_p - \vec{R}_q$.

With the help of Graf's addition theorem, we have

$$H_{m''}^{(1)}(k_0 r_q) e^{im''\theta_q} = \sum_{m'} g_{m'm''} J_{m'}(k_0 r_p) e^{im'\theta_p}, \tag{A.6}$$

where

$$g_{m'm''} = H_{m'-m''}^{(1)}(k_0 R_q) e^{i(m''-m')\Theta_q}. \tag{A.7}$$

Substituting Eqs. (A.6) and (A.7) into Eq. (A.5), we obtain

$$E_z^{in}(\vec{r}_p) = \sum_{q \neq p} \sum_{m"} \sum_{m'} \left( b_{m"}^q g_{m'm"} J_{m'}(k_0 r_p) e^{im'\theta_p} \right). \tag{A.8}$$

Combining Eq. (A.8) with Eq. (A.1) and using Eq. (A.3), we obtain the following self-consistent equation:

$$b_m^p = \sum_{m'} t_{mm'} \left( a_{m'}^p + \sum_{q \neq p} \sum_{m"} g_{m'm"} b_{m"}^q \right). \tag{A.9}$$

To obtain dispersion relations for a periodic array of scatterers, we drop the source terms, $a_{m'}^p$, and apply the Bloch theorem, $b_{m"}^q = b_{m"}^p e^{i\vec{K} \cdot \vec{R}_q}$, to Eq. (A.9), where $\vec{K}$ is the Bloch wave vector, yielding

$$\sum_m \left( \sum_{m'} t_{mm'} \sum_{q \neq p} g_{m'm"} e^{i\vec{K} \cdot \vec{R}_q} - \delta_{mm"} \right) b_m^p = 0. \tag{A.10}$$

Finally, the dispersion relation, $\omega(\vec{K})$, is obtained from the condition that gives nontrivial solutions to Eq. (A.10), i.e.,

$$\det \left| \sum_{m'} t_{mm'} \sum_{q \neq p} g_{m'm"} e^{i\vec{K} \cdot \vec{R}_q} - \delta_{mm"} \right| = 0. \tag{A.11}$$

**Appendix B  Evaluation of Lattice Sum for EM Waves**

To solve Eq. (A.11), one important step is to evaluated the lattice sum $\sum_{q \neq p} g_{m'm"} e^{i\vec{K} \cdot \vec{R}_q}$. Here we used the method proposed by Chin *et al* in Ref. [22]. We first expand the Hankel function in the momentum space as:

$$\sum_q H_0^{(1)} \left( k_0 | \vec{r} - \vec{R}_q | \right) e^{i\vec{K} \cdot \vec{R}_q} = \frac{4i}{\Omega} \sum_{h,n} \frac{i^n}{k_0^2 - Q_h^2} J_n(Q_h r) e^{-in\varphi_h} e^{in\theta}, \tag{B.1}$$

where $\Omega$ is the volume of the unit cell and $(Q_h, \phi_h)$ stands for vector $\vec{Q}_h = \vec{K} + \vec{K}_h$ in polar coordinates, and $\vec{K}_h$ is the reciprocal lattice vector. Alternatively, by using Graf's addition theorem, we can also express the Hankel function as

$$\sum_q H_0^{(1)}\left(k_0 \mid \vec{r} - \vec{R}_q \mid\right) e^{i\vec{K}\cdot\vec{R}_q}$$
$$= \sum_{q\neq p}\sum_n \left( H_n^{(1)}\left(k_0 R_q\right) e^{-in\Theta_q} J_n(k_0 r) e^{i\vec{K}\cdot\vec{R}_q} e^{in\theta} + H_0^{(1)}(k_0 r)\delta_{n0} e^{in\theta}\right). \tag{B.2}$$

By comparing Eq. (B.1) and (B.2), we obtain the relation

$$\sum_{q\neq p} H_n^{(1)}\left(k_0 R_q\right) e^{-in\Theta_q} e^{i\vec{K}\cdot\vec{R}_q} J_n(k_0 r) = \frac{4i^{n+1}}{\Omega} \sum_h \frac{J_n(Q_h r)}{k_0^2 - Q_h^2} e^{-in\phi_h} - H_0^{(1)}(k_0 r)\delta_{n0}. \tag{B.3}$$

This equation is satisfied as long as the condition of Graf's addition theorem holds, i.e., $r < \min R_q = a$. In order to improve computation precision, we multiply both sides by $r^{m+1}$ and integrate from 0 to $a$, yielding

$$\sum_{q\neq p} H_n^{(1)}\left(k_0 R_q\right) e^{-in\Theta_q} e^{i\vec{K}\cdot\vec{R}_q} J_{n+1}(k_0 a)$$
$$= \frac{4i^{n+1}k_0}{\Omega} \sum_h \frac{J_{n+1}(Q_h a)}{Q_h(k_0^2 - Q_h^2)} e^{-in\phi_h} - \left(H_1^{(1)}(k_0 a) + \frac{2i}{\pi k_0 a}\right)\delta_{n,0}. \tag{B.4}$$

Comparing Eq. (B.4) and Eq. (A.7), we get:

$$\sum_{q\neq p} g_{m'm''} e^{i\vec{K}\cdot\vec{R}_q} = \sum_{q\neq p} H_{m'-m''}^{(1)}(k_0 R_q) e^{i(m''-m')\Theta_q} e^{i\vec{K}\cdot\vec{R}_q}$$
$$= \frac{1}{J_{m'-m''+1}(k_0 a)} \left( \frac{4i^{m'-m''+1}k_0}{\Omega} \sum_h \frac{J_{m'-m''+1}(Q_h a)}{Q_h(k_0^2 - Q_h^2)} e^{-i(m'-m'')\phi_h} - \left(H_1^{(1)}(k_0 a) + \frac{2i}{\pi k_0 a}\right)\delta_{m'-m'',0}\right), \tag{B.5}$$

which shows that the lattice sum $\sum_{q\neq p} g_{m'm''} e^{i\vec{K}\cdot\vec{R}_q}$ is a function of the difference between two angular momentum quantum numbers, i.e., $(m' - m'')$, so that the lattice sum can also be expressed by:

$$\sum_{q\neq p} g_{m'm''} e^{i\vec{K}\cdot\vec{R}_q} = S(m' - m''), \tag{B.6}$$

where

$$S(n) = \frac{1}{J_{n+1}(k_0 a)} \left( \frac{4i^{n+1}k_0}{\Omega} \sum_h \frac{J_{n+1}(Q_h a)}{Q_h(k_0^2 - Q_h^2)} e^{-in\phi_h} - \left(H_1^{(1)}(k_0 a) + \frac{2i}{\pi k_0 a}\right)\delta_{n,0}\right) \quad (n\geq 0). \tag{B.7}$$
$$S(-n) = -S^*(n)$$

**Appendix C  Multiple-Scattering Theory and Lattice Sum for Elastic Waves**

The 2D elastic wave equation considers the displacement field, $\vec{u}$, in the *xy*-plane, which can be described by [19, 29]:

$$\rho(\vec{r})\frac{\partial u_i^2(\vec{r})}{\partial t^2} = \nabla \cdot \left(\mu(\vec{r})\nabla u_i(\vec{r})\right) + \nabla \cdot \left(\mu(\vec{r})\frac{\partial \vec{u}(\vec{r})}{\partial x_i}\right) + \frac{\partial}{\partial x_i}\left[\lambda(\vec{r})\nabla \cdot \vec{u}(\vec{r})\right]. \quad (C.1)$$

In general, $\vec{u}$ can be decoupled into a longitudinal part and a transverse part, i.e., $\vec{u} = \nabla \phi_l + \nabla \times (\phi_t \hat{e}_z)$, where $\phi_l$ and $\phi_t$ are the longitudinal and transverse gauge potentials, respectively. In the frequency domain, the solutions to $\phi_l$ and $\phi_t$ can also be expanded by using Bessel and Hankel functions. The wave incident on and scattered by a single scatterer can be written as:

$$\vec{u}_p^{in}(\vec{r}_p) = \sum_m \left( a_{lm}^p \nabla \left[ J_m(k_{l0}r_p)e^{im\theta_p} \right] + a_{tm}^p \nabla \times \left[ \hat{z} J_m(k_{t0}r_p)e^{im\theta_p} \right] \right), \quad (C.2)$$

and

$$\vec{u}_p^{sca}(\vec{r}_p) = \sum_m \left( b_{lm}^p \nabla \left[ H_m^{(1)}(k_{l0}r_p)e^{im\theta_p} \right] + b_{tm}^p \nabla \times \left[ \hat{z} H_m^{(1)}(k_{t0}r_p)e^{im\theta_p} \right] \right), \quad (C.3)$$

which are elastic analogues to Eqs. (A.1) and (A.2), respectively. $k_{l0} = \omega\sqrt{\rho_0/(\kappa_0 + \mu_0)}$ represents the longitudinal wave vector in the matrix and $k_{t0} = \omega\sqrt{\rho_0/\mu_0}$ denotes the corresponding transverse wave vector. In the presence of an inclusion, the above two parts are coupled through boundary conditions, which give:

$$\begin{pmatrix} \vec{b}_l^p \\ \vec{b}_t^p \end{pmatrix} = \begin{pmatrix} T^{ll} & T^{lt} \\ T^{tl} & T^{tt} \end{pmatrix} \begin{pmatrix} \vec{a}_l^p \\ \vec{a}_t^p \end{pmatrix}, \quad (C.4)$$

where $\vec{b}_\alpha^p$ and $\vec{a}_\alpha^p$ ($\alpha = l,t$) are $1 \times m$ column vectors, $T^{\alpha\beta}$ are $m \times m$ matrices with matrix elements $t_{mm'}^{\alpha\beta} = D_m^{\alpha\beta}\delta_{mm'}$. $D_m^{\alpha\beta}$ are the elastic Mie-like scattering coefficients of the inclusion, whose explicit expressions are very lengthy and can be found in the Appendix A of Ref. [19].

To obtain the dispersion relations for elastic metamaterials, we adopt the similar method to the EM case. The MST for an elastic metamaterial with periodic lattice structure gives the following secular equation :

$$\det \left| \begin{pmatrix} T^{ll}G^l & T^{lt}G^t \\ T^{tl}G^l & T^{tt}G^t \end{pmatrix} - I \right| = 0. \tag{C.5}$$

Here $G^l$ and $G^t$ $m \times m$ matrices, whose elements are given by the lattice sum $G^{\beta}_{m'm''} = \sum_{q \neq p} g^{\beta}_{m'm''} e^{i\vec{K} \cdot \vec{R}_q}$ ($\beta = l, t$), which satisfies:

$$\sum_{q \neq p} g^{\beta}_{m'm''} e^{i\vec{K} \cdot \vec{R}_q} = S(\beta, m' - m''), \tag{C.6}$$

where

$$S(\beta, n) = \frac{1}{J_{n+1}(k_{\beta 0}a)} \left( \frac{4i^{n+1}k_{\beta 0}}{\Omega} \sum_h \frac{J_{n+1}(Q_h a)}{Q_h(k_{\beta 0}^2 - Q_h^2)} e^{-in\phi_h} - \left( H_1^{(1)}(k_{\beta 0}a) + \frac{2i}{\pi k_{\beta 0}a} \right) \delta_{n,0} \right) \quad n \geq 0. \tag{C.7}$$

$$S(\beta, -n) = -S^*(\beta, n).$$

$$\tilde{\varepsilon}_s = 2F(k_s r_s)\varepsilon_s \Big/ \Big[1 + k_s^2 r_s^2 \ln(k_0 r_s) F(k_s r_s)\mu_0/\mu_s\Big].$$

**Figures**

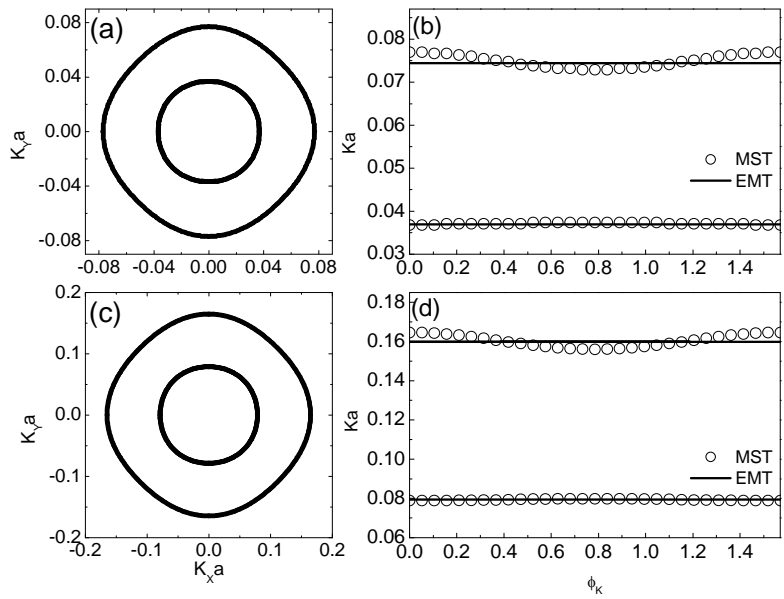

Fig. 1 (a) The EFS of a square lattice of silicone rubber cylinders embedded in an epoxy host at $\tilde{f}_1 = 0.01$. The rubber cylinder's radius is 0.2 lattice constant. (b) $Ka$ as a function of $\phi_K$. The open circles are calculated by the MST method while the solid lines are predicted by EMT. (c) and (d) are the same as (a) and (b), respectively, where the frequency is $\tilde{f}_2 = 0.03$.

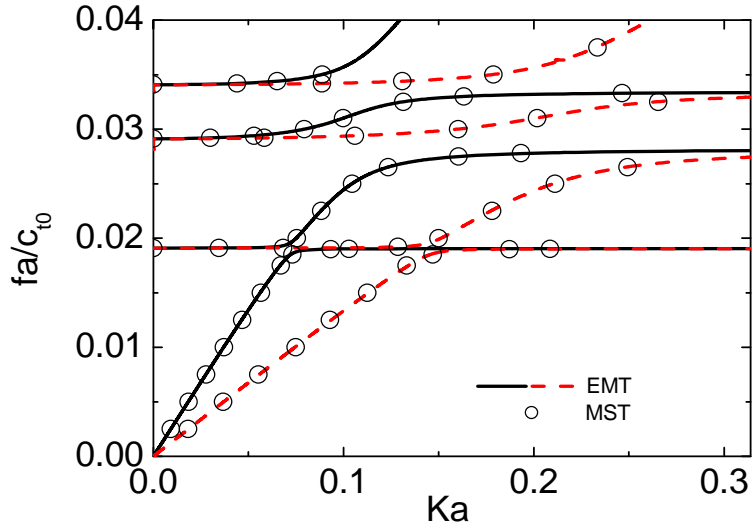

Fig. 2 (Color online) Open circles are angle-averaged dispersion relations calculated using MST for a square array of silicone rubber cylinders embedded in an epoxy host. The cylinder has a radius of 0.2 lattice constant. The black solid and red dashed curves are, respectively, the quasi-longitudinal and quasi-transverse branches obtained from EMT. Dimensionless frequencies, $fa/c_{t0} = \omega a/(2\pi c_{t0})$, are used, where $c_{t0}$ is the transverse wave speed in the host.